\documentstyle[12pt,epsf]{article}
\begin{document}
\begin{center}
{\Large \bf The Role of Gluons in Dilepton Production
from the Quark-Gluon Plasma}\\[2ex]
{Ziwei Lin and C. M. Ko}\\[2ex]
{Cyclotron Institute and Physics Department, Texas A\&M University,
College Station, TX 77843}\\[2ex]
\end{center}

\begin{abstract}
We study high mass dilepton production from gluon-induced processes, $g q
\rightarrow q \gamma^*$, $g \bar q \rightarrow \bar q \gamma^*$,
and $g g \rightarrow q \bar q \gamma^*$, in a thermally
equilibrated but chemically under-saturated partonic matter that is
expected to be created in the initial stage of ultra-relativistic
heavy ion collisions.  Regulating the divergence in these
processes by the thermal quark mass, we find that gluon-induced
processes are more important than the leading-order $q\bar q$
annihilation process as a result of the larger number of gluons
than quarks in the partonic matter. The dependence of the
thermal dilepton yield from the partonic stage of heavy ion
collisions on the initial conditions for the partonic matter 
is also studied. We further
discuss the feasibility of observing thermal dileptons from the
quark-gluon plasma in heavy ion experiments.

\end{abstract}

\section{Introduction}

One of the signatures for the formation of a
quark-gluon plasma (QGP) in heavy ion collisions at the
Relativistic Heavy Ion Collider (RHIC) is the thermal dileptons it
produces. Many studies have been carried out to estimate the
dilepton yield from the QGP. In earlier works, 
contributions from the leading-order (LO) $q\bar q$ annihilation
process \cite{keijo} and processes up to ${\cal O}(\alpha_s)$ 
\cite{baier,cleymans} have been evaluated for a thermally and 
chemically equilibrated QGP.
Although thermal
equilibrium can be reached in the partonic matter created in heavy
ion collisions at RHIC energies, chemical equilibrium is, however,
unlikely to be established \cite{hotglue,biro}. 
Later studies have thus included the effects of chemical non-equilibrium
on thermal dilepton and photon production from QGP 
\cite{shuryak,kampfer,strickland,traxler}.

Since the partonic matter is dominated by gluons, the thermal dilepton
yield from the LO $q\bar q$ annihilation process is thus much
smaller than that from the initial Drell-Yan process and final
charm decays \cite{xn-pr}. However, the dominance of gluons 
in the initial partonic matter provides the
interesting possibility that dilepton production from
gluon-induced processes may become more important than the LO $q\bar
q$ annihilation process, even though they are suppressed by
powers of $\alpha_s$. In this paper, we shall study high mass thermal
dilepton production from gluon-induced processes, i.e., the
Compton processes involving one gluon in the initial state 
($g q \rightarrow q \gamma^*$ and $g \bar q
\rightarrow \bar q \gamma^*$) and the two-gluon fusion process
($g g\rightarrow q \bar q \gamma^*$). In particular, we 
shall consider these processes only at leading-order in $\alpha_s$, 
i.e., at ${\cal O}(\alpha_s)$ and ${\cal O}(\alpha_s^2)$ for the
Compton and two-gluon fusion processes, respectively.

This paper is organized as follows. In Section \ref{sec-thqm}, we
introduce the thermal quark mass in a non-equilibrium partonic
matter at finite temperature, which will be used in Section
\ref{sec-xsection} to regularize the divergent dilepton production
cross sections for gluon-induced reactions. Thermal dilepton rates
are then calculated in Section \ref{sec-rate} for an
under-saturated partonic matter in thermal equilibrium. In Section
\ref{sec-rhic}, we estimate the total thermal dilepton yield in
heavy ion collisions at RHIC energies and study its dependence on
the initial conditions for the partonic matter. We then compare it 
with the Drell-Yan yield to see if thermal dileptons from the QGP 
can be observed in experiments. Finally, a summary and discussions 
are given in Section \ref{sec-final}.

\section{The Thermal Quark Mass}
\label{sec-thqm}

For a partonic matter in thermal equilibrium at temperature $T$
but out of chemical equilibrium, the parton distribution can be
written in the J\"uttner form,
\begin{eqnarray} 
f_i=\frac {\lambda_i}{e^{\beta
u\cdot p} \pm \lambda_i}, \label{juttner}
\end{eqnarray}
where $i$ denotes different parton species, $\beta=1/T$, and $u$
is the four velocity of the local comoving frame. In the limit of
small fugacities ($\lambda_i \ll 1$), the parton distribution
functions reduce to the Boltzmann form, 
\begin{eqnarray} 
f_i\simeq \lambda_i e^{-\beta u\cdot p}. \label{bf} 
\end{eqnarray} 
Assuming that the partonic matter is both baryon and flavor 
symmetric with $N_f$ quark flavors, i.e., 
$\lambda_{q_i}=\lambda_{\bar q_i}=\lambda_q$ ($i=1,N_f$), then 
two fugacity parameters, $\lambda_g$ and $\lambda_q$, are needed 
to specify the degree of chemical equilibrium.

Since chemical equilibration of the partonic matter created in
heavy ion collisions at RHIC is likely to be slow, we shall use
the Boltzmann distribution of Eq.(\ref{bf}) in the following
calculations unless specified otherwise. In this case, the parton
density in the local comoving frame is directly related to the
fugacity, 
\begin{eqnarray} 
n_g=\lambda_g \frac {d_g T^3}{\pi^2}, \qquad
n_q+n_{\bar q}=\lambda_q \frac {\left (d_q+d_{\bar q}\right )N_f
T^3}{\pi^2},
\end{eqnarray}
if partons are taken to be massless. In the above, $d_i$ is the
degeneracy factor of parton $i$, i.e., $d_g=2(N_c^2-1)$ and
$d_q=d_{\bar q}=2N_c$.

The thermal quark mass, denoted as $\mu$, is given by
\begin{eqnarray}
\mu^2&=& 4\pi \alpha_s \frac {N_c^2-1}{2N_c}
\int \frac {d^3p}{(2\pi)^3} \frac{1}{p} (f_g + f_q),
\end{eqnarray}
where $\alpha_s=g^2/(4\pi)$ is the QCD strong coupling. For a QGP
in chemical equilibrium, this gives a thermal quark mass
$\mu^2=g^2 T^2/6$ when using the exact parton distribution of
Eq.(\ref{juttner}). For an under-saturated partonic matter
described approximately by Eq.(\ref{bf}), the thermal quark mass
becomes
\begin{eqnarray}
\mu^2=\frac{2g^2 T^2}{3\pi^2} ( \lambda_g + \lambda_q).
\end{eqnarray}

\section{Dilepton Production Cross Sections}
\label{sec-xsection}

\begin{figure}[ht]
\setlength{\epsfxsize=0.6\textwidth}
\setlength{\epsfysize=0.75\textheight}
\centerline{\epsfbox{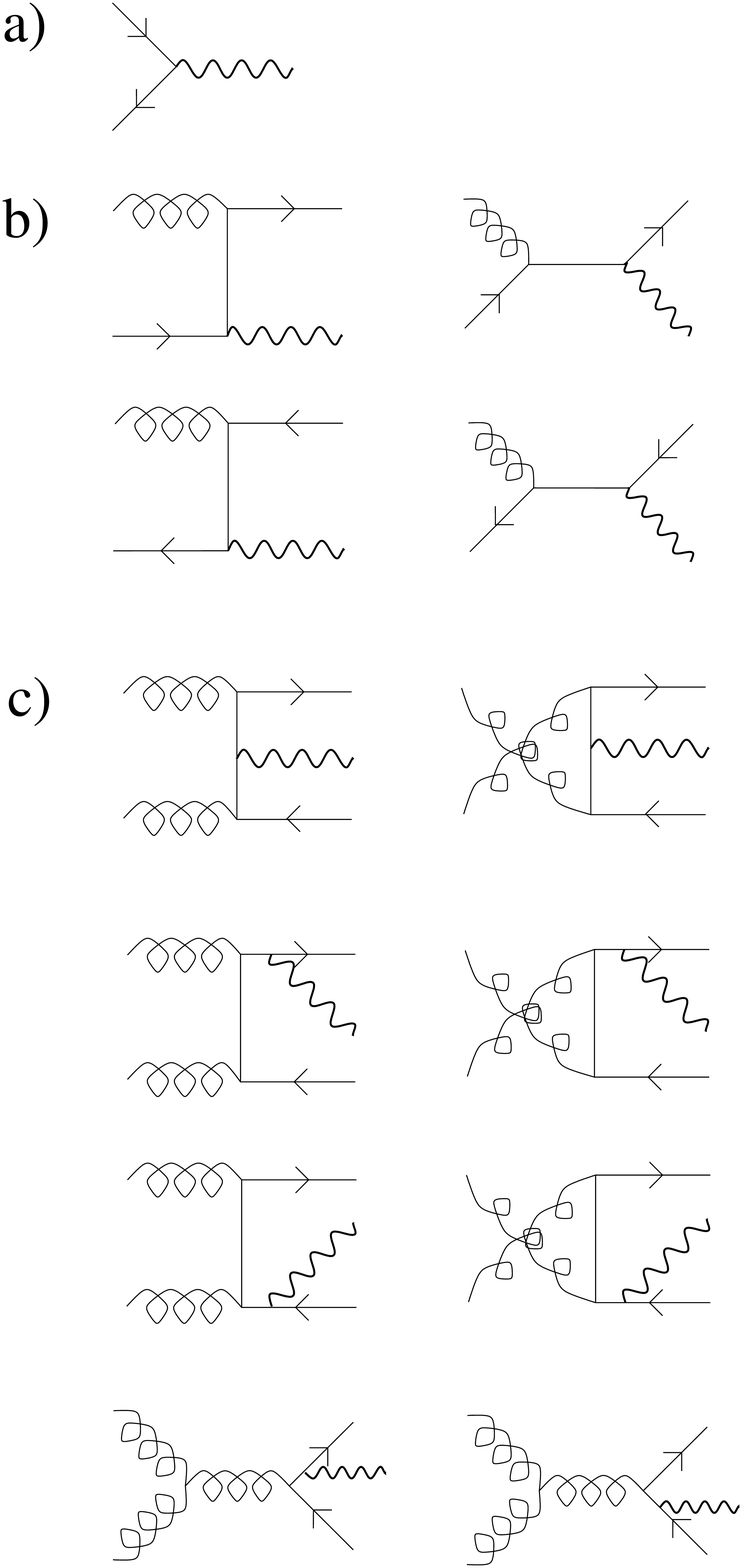}}
\caption{
Diagrams for dilepton production: a) $q\bar q$ annihilation,  b) the Compton 
processes, and c) two-gluon fusion processes. Curly lines denote 
gluons, wavy lines denote virtual photons, and straight lines with 
arrows to the right (left) denote quarks (antiquarks).
}
\label{fig_diagrams}
\end{figure}

Leading-order processes in the strong coupling constant $\alpha_s$ 
for dilepton production from parton-parton interactions 
are $q\bar q$ annihilation ($q\bar q \rightarrow \gamma^*$), 
Compton processes ($gq \rightarrow q \gamma^*$ and $g\bar q \rightarrow 
\bar q\gamma^*$), two-gluon fusion ($gg \rightarrow q \bar q \gamma^*$),
and two-quark ($qq \rightarrow qq \gamma^*$) and two-antiquark
bremsstrahlung ($\bar q \bar q \rightarrow \bar q \bar q
\gamma^*$). They are of order ${\cal O}(1), {\cal O}(\alpha_s),
{\cal O}(\alpha_s^2), {\cal O}(\alpha_s^2)$ and ${\cal
O}(\alpha_s^2)$, respectively. In the present study, 
we neglect two-quark and two-antiquark
bremsstrahlung as they do not involve gluons in the initial state
and are also suppressed by $\alpha_s^2$ relative to the LO $q\bar q$ 
annihilation process. Furthermore, we neglect $3 \rightarrow 2$ 
processes as they are expected to be less important when parton 
fugacities are much below one. The processes included in our study 
for dilepton production are shown in Fig.\ref{fig_diagrams}.

The cross section for dilepton production from partons can be generally
written as
\begin{eqnarray}
\frac{d\hat \sigma_{ij}}{dM^2} = \frac{4\pi \alpha^2}{9M^2 \hat s} W_{ij}.
\end{eqnarray}
In the above, indices $i$ and $j$ specify initial-state partons, 
$\hat s= (p_i+p_j)^2$, and $\alpha$ is the fine structure constant. 
The quantity $W_{ij}$ is related to the square of the transition 
amplitude. For the leading-order $q\bar q$ annihilation process, 
it is simply
\begin{eqnarray}
W_{q\bar q}=e_q^2 \delta(1-\xi),
\end{eqnarray}
with $\xi=M^2/\hat s$.

For the Compton processes, the squared matrix element, after averaging
over initial and summing over final spins and colors, is given by
\begin{eqnarray} 
\overline {|M_{gq\rightarrow q\gamma^*}|^2} &=& \overline
{|M_{g\bar q\rightarrow \bar q \gamma^*}|^2} =\frac {(4\pi)^2
\alpha \alpha_s e_q^2}{3} \left [-\frac{\hat t}{\hat s}-\frac{\hat
s}{\hat t} +\frac {2M^2(\hat s+\hat t-M^2)}{\hat s\hat t} \right
]. \label{gq}
\end{eqnarray}

The total cross section, given by $\hat \sigma= \int d\hat t
\overline {|M|^2}/(16\pi \hat s^2)$, is divergent. 
Following previous studies of low mass dilepton and photon production
from the QGP \cite{altherr,thoma,baier2}, we use
the thermal quark mass as the screening mass for regularizing 
all divergence in the cross sections. 
We thus replace $1/\hat t$ with $1/(\hat t-\mu^2)$, 
and Eq.(\ref{gq}) then becomes
\begin{eqnarray}
\overline {|M_{gq\rightarrow q\gamma^*}|^2} &\simeq&\frac
{(4\pi)^2 \alpha \alpha_s e_q^2}{3} \left [-\frac{\hat t}{\hat
s}-\frac{\hat s}{\hat t-\mu^2} +\frac {2M^2(\hat s+\hat
t-M^2)}{\hat s(\hat t-\mu^2)} \right ].
\end{eqnarray}
After integrating over $\hat t$, we obtain
\begin{eqnarray} 
W_{g q}=W_{g\bar q}=\frac{\alpha_s e_q^2}{4\pi} c_0(\xi), \label{wgq}
\end{eqnarray}
where
\begin{eqnarray}\label{screen1} 
c_0(\xi)=\frac{1+2\xi-3\xi^2}{2}+ \left[ 1-2\xi \left
(1-\xi+\frac{\mu^2}{M^2}\xi \right ) \right ] \ln \left
(1+\frac{M^2}{\mu^2}\frac{1-\xi}{\xi} \right ).
\end{eqnarray}

There are no studies on dilepton production from the two-gluon fusion
process, $gg\rightarrow q\bar q \gamma^*$, at finite
temperature. However, processes involving heavy quarks in the
final state such as $gg\rightarrow Q\bar Q \gamma^*$ have been
studied in determining the K-factor in Drell-Yan processes in the
limit of $M^2 \gg m_Q^2$ \cite{gg,dyk}. These results can be adopted 
for the production of dileptons with invariant mass $M \geq 2$ GeV.
Since $M^2 \gg \mu^2$ unless the temperature is extremely high, we
can replace the heavy quark mass $m_Q$ in Ref. \cite{gg,dyk} with
the thermal quark mass $\mu$ and obtain
\begin{eqnarray}\label{screen2}
W_{gg}=e_q^2 \left (
\frac{1}{2}\right )^2 \left ( \frac{\alpha_s}{4\pi}\right )^2
\left [ c_1(\xi) \ln^2 \frac{M^2}{\mu^2} + c_2(\xi) \ln
\frac{M^2}{\mu^2} +c_3(\xi) \right ]. \label{wgg}
\end{eqnarray}
In the above, the functions $c_1(\xi), c_2(\xi)$ and $c_3(\xi)$ 
can be obtained from Ref.\cite{gg}, where the calculation was 
carried out in the $\overline {\rm MS}$ scheme.

\section{Thermal Dilepton Rates}
\label{sec-rate}

For a partonic matter in thermal equilibrium at temperature $T$
but out of chemical equilibrium, the thermal dilepton rate is 
given by
\begin{eqnarray} 
\frac{dR_{ij}}{dM}&=& S_{ij} \sum \int \frac{d^3
p_1}{(2\pi)^3}f_i(p_1)\frac{d^3 p_2}{(2\pi)^3}f_j(p_2) v_{\rm rel}
\frac{d\hat \sigma_{ij}}{dM}(M,\sqrt {\hat s}), \label{rate}
\end{eqnarray}
with $v_{\rm rel}=(p_1\cdot p_2)/(p_1 p_2)$, $S_{gg}=1/2$
accounting for the two identical gluons in the initial state, and
$S_{ij}=1$ otherwise. The summation is over the parton spin and color
indices.

The integrals in the above equation can be evaluated using the 
rapidity and transverse momentum variables, 
\begin{eqnarray} 
&&\hat s=M_0^2,~~ P_0^\mu= \left (M_{\perp 0}
\cosh Y_0, P_{\perp 0} \cos \Phi_0, P_{\perp 0} \sin \Phi_0,
M_{\perp 0} \sinh Y_0 \right ), \nonumber \\
&&p_1^\mu= p_{\perp 1} \left ( 
\cosh y_1, \cos \phi_1, \sin \phi_1, \sinh y_1 \right ),
\label{rapidity}
\end{eqnarray}
where $M_{\perp 0}=\sqrt {M_0^2+P_{\perp 0}^2}$.  
Inserting $\int d^4 P_0 \delta^4 (P_0-p_1-p_2)$ into
Eq.(\ref{rate}), the delta function can then be used
to remove the $\int d^3 p_2$ and $\int d p_{\perp 1}$ integrals.
After integrating over $\Phi_0$ and redefining $y_1-Y_0$ and $\phi_1-
\Phi_0$ as $y_1$ and $\phi_1$, we obtain
\begin{eqnarray}
\frac{dR_{ij}}{dM}&=& \frac{d_id_j \lambda_i
\lambda_j S_{ij}}{4(2\pi)^5} \int dM_0 dY_0 dM_{\perp 0} d\phi_1 d
y_1 \nonumber \\ && \times \frac{ M_{\perp 0} M_0^5 e^{-M_{\perp
0} \cosh (Y_0-\eta)/T}} {\left ( M_{\perp 0}\cosh y_1-P_{\perp
0}\cos \phi_1 \right )^2} \frac{d\hat \sigma_{ij}}{dM}(M,M_0).
\label{rate1}
\end{eqnarray}
Carrying out the integrals over $Y_0$ and $\phi_1$ and then
those over $y_1$ and $M_{\perp 0}$ gives
\begin{eqnarray}
\frac{dR_{ij}}{dM}&=& \frac{d_id_j \lambda_i \lambda_j
S_{ij}}{(2\pi)^4} \int_M dM_0 M_0^4 T K_1(M_0/T) \frac{d\hat
\sigma_{ij}}{dM}(M,M_0).
\end{eqnarray}
Summing over quark flavors, we finally have
\begin{eqnarray}
\frac{dR_{q\bar
q}}{dM}&=& \alpha^2 \lambda_q^2 \frac{F_q}{\pi^3} M^2 T K_1(M/T) ,
\label{rate2}
\\ 
\frac{dR_{gq+g\bar q}}{dM}&=& \alpha^2 \alpha_s
\lambda_g \lambda_q \frac{8 F_q}{3\pi^4}\frac{T}{M} \int_M dM_0
M_0^2 K_1(M_0/T) c_0(\xi), 
\\ 
\frac{dR_{gg}}{dM}&=&
\alpha^2 \alpha_s^2 \lambda_g^2 \frac{F_q}{9\pi^5} \frac{T}{M}
\int_M dM_0 M_0^2 K_1(M_0/T) \nonumber \\ && \times \left [
c_1(\xi) \ln^2 \frac{M^2}{\mu^2} + c_2(\xi) \ln \frac{M^2}{\mu^2}
+c_3(\xi) \right ], 
\end{eqnarray}
with
\begin{eqnarray}
F_q=\sum_{i=1}^{N_f} e_{q_i}^2=\frac {5}{9} {\;\;\rm for\;\;}N_f=2.
\end{eqnarray}

For the strong coupling constant, we use the LO value
with three quark flavors,
\begin{eqnarray}
\alpha_s = \frac{4\pi}{9 \ln \frac{M^2}{\Lambda^2}},
\end{eqnarray}
with  $\Lambda=200$ MeV and the normalization scale chosen to be
the dilepton mass $M$.

In the limit $M^2 \gg \mu^2$, the thermal dilepton rates depend on
the coupling constant and fugacities as
\begin{eqnarray}
\frac{dR_{q\bar q}}{dM}& \propto &  \alpha^2 \lambda_q^2, \nonumber \\
\frac{dR_{gq+g\bar q}}{dM}&\propto& \alpha^2\alpha_s \lambda_g \lambda_q
\ln \frac {1}{\lambda_g+\lambda_q}, \nonumber \\
\frac{dR_{gg}}{dM}& \propto & \alpha^2 \alpha_s^2 \lambda_g^2
\ln^2 \frac{1}{\lambda_g+\lambda_q}.
\label{scale}
\end{eqnarray}

\begin{figure}[ht]
\setlength{\epsfxsize=\textwidth}
\setlength{\epsfysize=0.7\textheight}
\centerline{\epsfbox{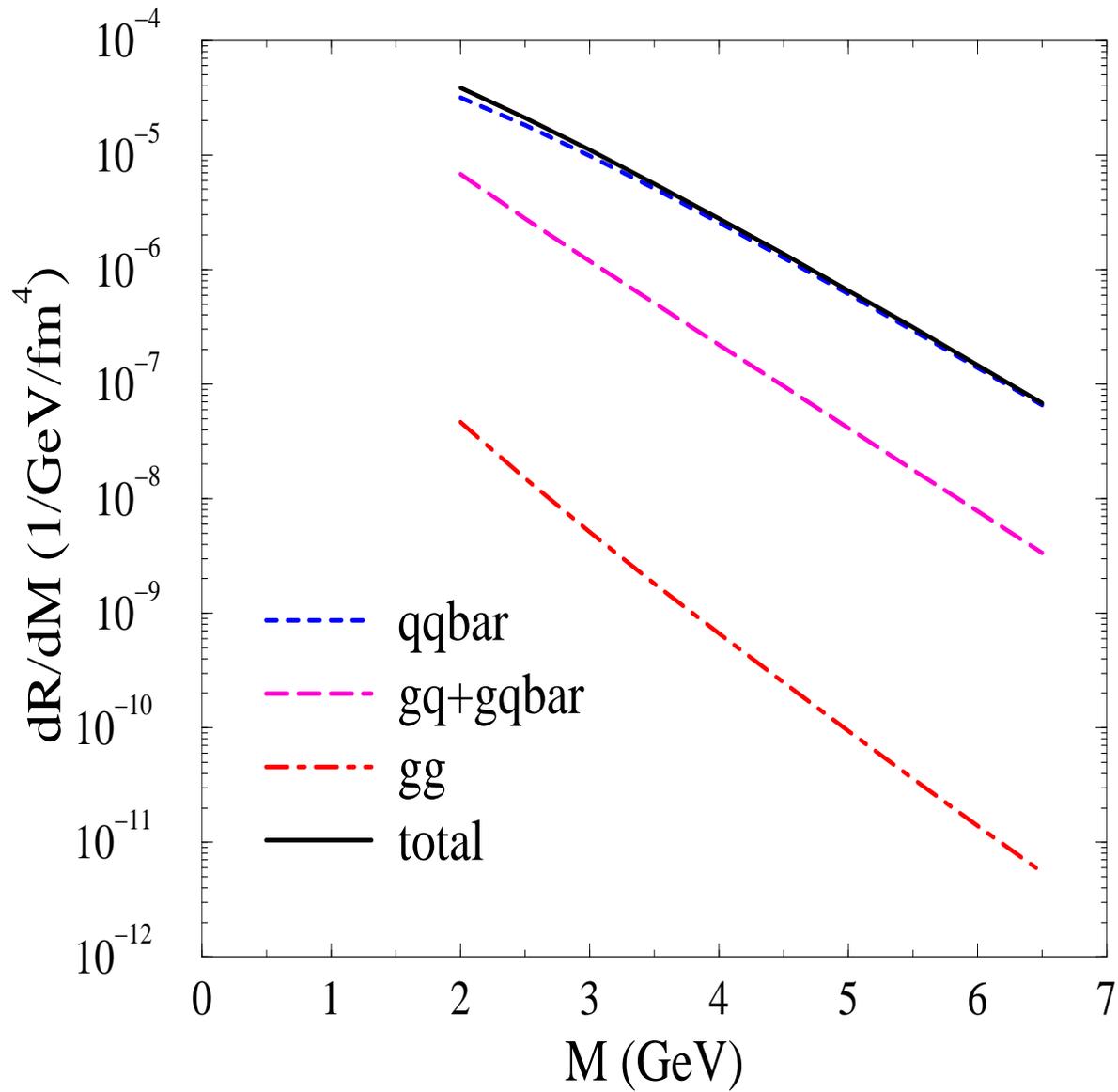}}
\caption{
Thermal dilepton rates from a quark-gluon plasma
in full thermal and chemical equilibrium at a temperature $T=570$ MeV.
}
\label{fig_full}
\end{figure}

In Fig. \ref{fig_full}, we show the thermal dilepton rates from a
quark-gluon plasma in full thermal and chemical equilibrium at
$T=570$ MeV. It is seen that for dileptons with mass above 2 GeV
the LO $q\bar q$ annihilation process gives the dominant
contribution, while the Compton processes give a small contribution,
and that from two-gluon fusion process is even smaller. These results
are not unexpected as the Compton and two-gluon fusion processes are
suppressed by $\alpha_s$ and $\alpha_s^2$, respectively.

\begin{figure}[ht]
\setlength{\epsfxsize=\textwidth}
\setlength{\epsfysize=0.7\textheight}
\centerline{\epsfbox{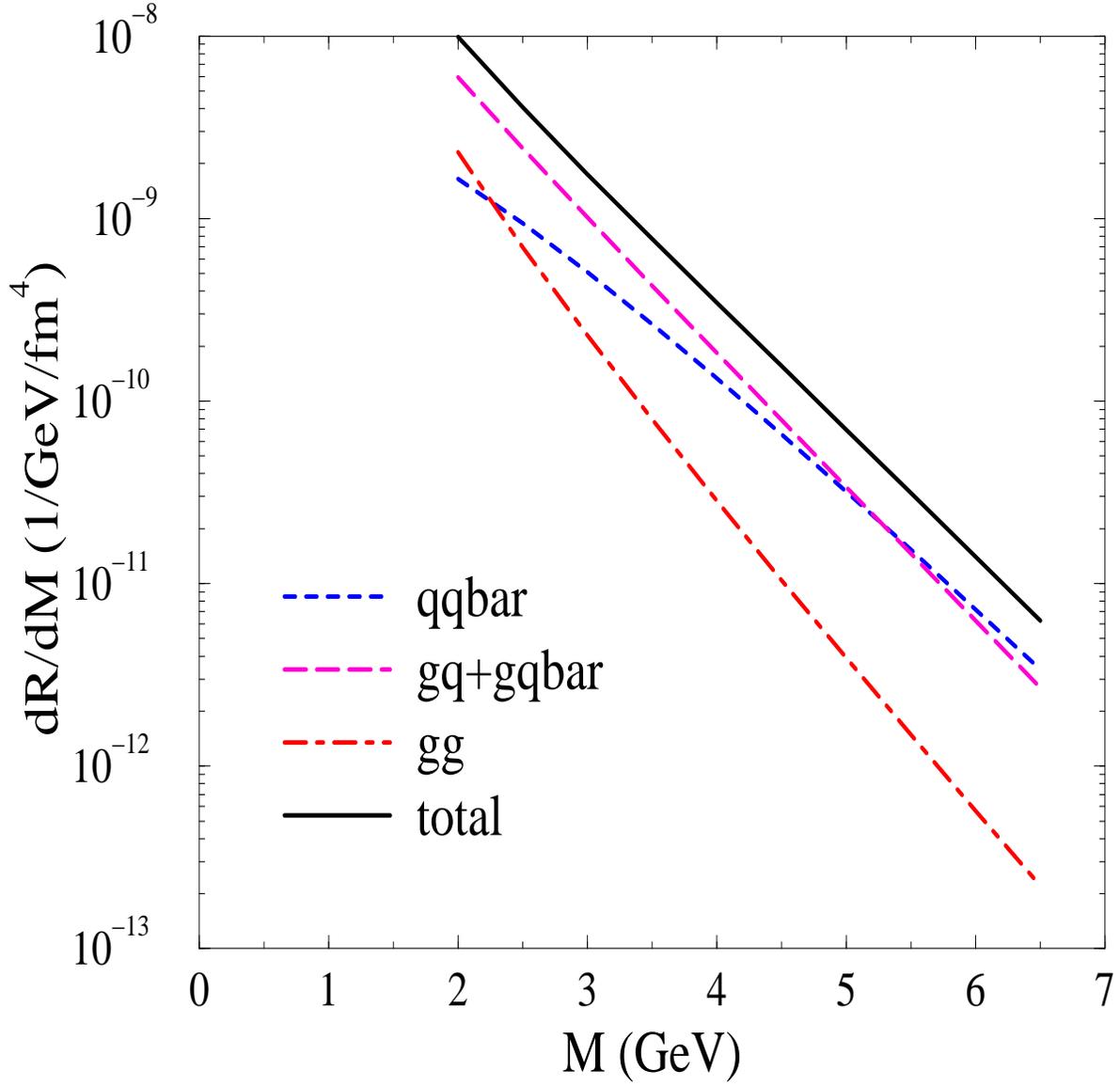}}
\caption{
Thermal dilepton rates from a gluon-dominated, under-saturated
quark-gluon plasma with $T=570$ MeV, $\lambda_g=0.060$ and $\lambda_q=0.0072$.
}
\label{fig_rate}
\end{figure}

The situation is, however, very different if partons are not in
chemical equilibrium. As an example, we show in Fig.
\ref{fig_rate} the thermal dilepton rates for $T=570$ MeV,
$\lambda_g=0.060$ and $\lambda_q=0.0072$. The reason for choosing
these values will be given in the next Section. In this case,
the Compton processes produce more thermal dileptons than
the LO $q\bar q$ annihilation, and even two-gluon fusion
processes give a comparable contribution to lower-mass dileptons.
As a result, the total thermal dilepton yield is significantly
enhanced by gluonic processes.

The appreciable difference between the results shown in Fig.
\ref{fig_rate} and Fig. \ref{fig_full} can be qualitatively
understood from the scaling behavior given in Eq.(\ref{scale}).
Because of the quadratic dependence on parton fugacities, gluonic
processes become dominant over the $q\bar q$ process when gluons
greatly outnumber quarks ($\lambda_g/\lambda_q\gg 1$). Even if
$\lambda_g/\lambda_q$ is fixed but $\lambda_g \rightarrow 0$ in an
under-saturated partonic matter, both $R_{gq+g\bar q}/R_{q\bar q}$
and $R_{gg}/R_{q\bar q}$ become logarithmic divergent, indicating that gluonic
processes become more important than $q\bar q$ annihilation when 
the parton system is further away from chemical equilibrium.

\begin{figure}[ht]
\setlength{\epsfxsize=\textwidth}
\setlength{\epsfysize=0.7\textheight}
\centerline{\epsfbox{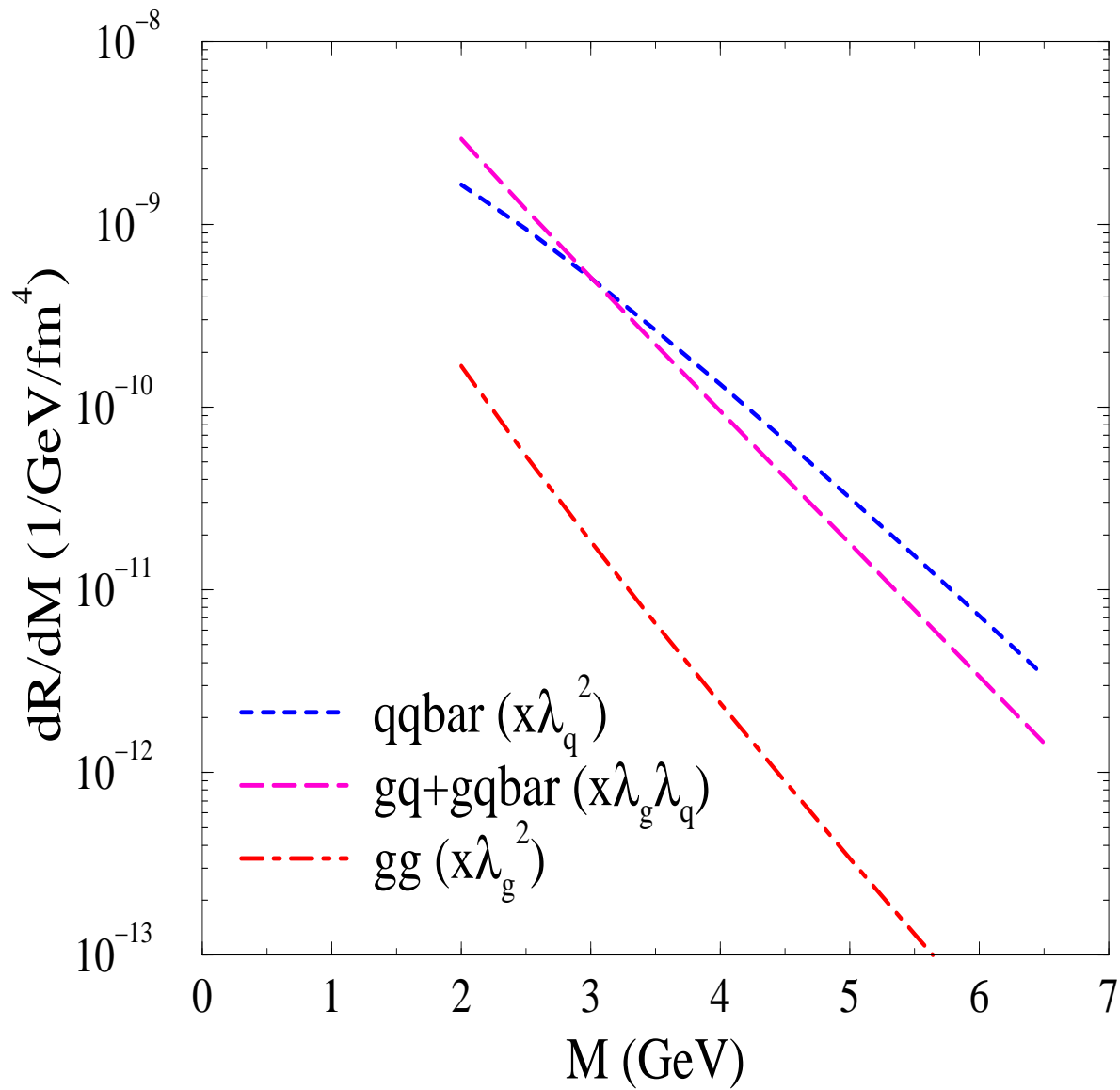}}
\caption{
Thermal dilepton rates from a quark-gluon plasma in
full thermal and chemical equilibrium at $T=570$ MeV
after being multiplied by the fugacity factors
$\lambda_g=0.060$ and $\lambda_q=0.0072$.
}
\label{fig_full2}
\end{figure}

To illustrate this quantitatively, we show in Fig. \ref{fig_full2}
the full-equilibrium results in Fig. \ref{fig_full} multiplied by
the quadratic fugacity factors. In this case, the Compton processes are 
already as important as the $q\bar q$ annihilation process. Further increase 
of the gluonic contribution when going from Fig. \ref{fig_full2} 
to Fig. \ref{fig_rate} is due to the logarithmic terms in 
Eq.(\ref{scale}), which come from the thermal quark mass in the 
regularized partonic cross sections given by Eqs.(\ref{wgq}-\ref{wgg}).

We also see in Figs. \ref{fig_full}-\ref{fig_full2} that the LO $q\bar 
q$ annihilation process is most effective in producing high 
invariant-mass dileptons because all incoming energy can be used. 
In contrast, there are one or two partons in the final state 
in the Compton and two-gluon fusion processes to carry away part of the 
incoming energy.  To produce high mass dileptons from these processes 
thus requires more energetic partons, which are less available.
Therefore, the three curves in Fig. \ref{fig_rate} corresponding
to the above three processes exhibit increasingly larger slopes.

\section{Thermal Dilepton Yield at RHIC}
\label{sec-rhic}

In this Section, we estimate the thermal dilepton yield in central
Au+Au collisions at RHIC by integrating the thermal dilepton rate
over the space-time four volume, which will be taken from a schematic 
model. Also, we will study the feasibility of observing the thermal
dilepton signal in experiments by comparing it with the Drell-Yan
background.

Starting from Eq.(\ref{rate1}) and using $d^4 x=\tau d\tau d\eta
d^2 x_\perp$, we first integrate over $\eta$ instead of $Y_0$ and
then over $\phi_1$, $y_1$ and $M_{\perp 0}$ as in previous
section. This gives
\begin{eqnarray}
\frac{dN_{ij}}{dM}&=& \int dY_0 d\tau
\tau d^2 x_\perp \frac{d_id_j \lambda_i \lambda_j
S_{ij}}{(2\pi)^4} \int_M dM_0 M_0^4 T K_1(M_0/T) \frac{d\hat
\sigma_{ij}}{dM}(M,M_0), 
\end{eqnarray}
which then leads to 
\begin{eqnarray}
\frac{dN_{ij}}{dM^2 dY_0} =\int_{\tau_0}^{\tau_f} d\tau \tau d^2
x_\perp \frac{dR_{ij}}{dM^2}(\tau). \label{dndm2dy}
\end{eqnarray}
In the above, $\tau_0$ is the initial proper time when the parton system 
first reaches thermal equilibrium, and $\tau_f$ is the final proper time 
when the temperature drops to $T_c$ (chosen to be $200$ MeV) and 
the partonic matter makes a transition to the hadronic matter. The 
rate $dR_{ij}/dM^2$ depends on $\tau$ implicitly through the time 
evolution of the variables $\lambda_i, T$ and $\mu$.

We note that while the pair rapidity $Y_0$ of the two partons in the
initial state is equal to the pair rapidity $Y$ of the dilepton
in the LO $q\bar q$ annihilation process, they are different in 
gluon-induced processes. However, if we assume that the expansion 
of the partonic matter follows the Bjorken scaling, the dependence of 
the thermal dilepton yield on $Y$ and $Y_0$ must have the following 
functional form:
\begin{eqnarray}
\frac{dN}{dM^2 dY dY_0} =
F(Y-Y_0,M),
\end{eqnarray} 
as a result of boost invariance of the integrand on the right-hand 
side of Eq.(\ref{rate}). This gives 
\begin{eqnarray} 
\frac{dN}{dM^2 dY} =\frac{dN}{dM^2 dY_0}.
\end{eqnarray}
The same argument applies to $\eta$, and it is easy to verify that
\begin{eqnarray}
\frac{dN}{dM^2 d\eta} =\frac{dN}{dM^2 dY_0}.
\end{eqnarray}

\subsection{Time evolution of parton fugacities and temperature}
\label{subsec-time}

First, we need to specify the time evolution of partons in the
quark-gluon phase created in heavy ion collisions at RHIC. Earlier
studies \cite{xn-pr} based on kinetic equations including
$2\leftrightarrow 3$ gluonic processes have shown that thermal
equilibrium is achieved first, and both gluon and quark fugacities
increase as the system evolves. However, since the quark-gluon plasma
has only a finite lifetime of a few fm/$c$, chemical equilibrium is 
unlikely to be established, and parton fugacities thus remain well 
below the value of one.  To take into account such effects, we
follow Fig. 25 of Ref.\cite{xn-pr} by simply assuming that the parton 
fugacities increase linearly with the proper time, 
\begin{eqnarray}
\lambda_i=\lambda_{i,0} \left ( \frac {\tau}{\tau_0}
\right ). \label{timef}
\end{eqnarray}

In an ideal hydrodynamic description, temperature of the
partonic matter decreases as $T \propto 1/\tau^{1/3}$, which then
leads to a decrease of the rapidity density of transverse energy 
as $dE_{\rm T}/dy \propto \tau T^4\propto 1/\tau^{1/3}$. In the 
parton cascade model, the time evolution of the rapidity density 
of transverse energy depends sensitively on the magnitude of parton 
cross sections and densities \cite{et}. Since we do not expect the 
rapidity density of transverse energy to change more than 25\% 
during the partonic stage of RHIC heavy ion collisions \cite{et,ampt}, 
we assume for simplicity that $dE_{\rm T}/dy$, given by
\begin{eqnarray}
\frac{dE_{\rm T}}{dy} = \frac{dN}{dy}\!\! <\!p_\perp \!\!>
\;\propto \; \left ( \sum_i d_i \lambda_{i,0} \right ) \tau T^3 T
\;\propto \; \left (\tau T^2 \right )^2, \label{detdy}
\end{eqnarray}
is a constant.
This then leads to the time evolution of temperature as
\begin{eqnarray}
T=T_0 \left (\frac {\tau_0}{\tau}\right )^{1/2}. \label{timet}
\end{eqnarray}
In the above, we have used 
\begin{eqnarray} 
\frac{dN}{dy} = \tau_0 \pi R_A^2\sum_i n_i \label{dndy} 
\end{eqnarray} 
at central rapidity at $\tau_0$ and $R_A\simeq 1.12 A^{1/3}$ fm. 
Also, we neglect the transverse expansion throughout this study.
The effect of transverse expansion on the production of dileptons 
of mass above $2$ GeV has been shown to be very small in heavy 
ion collisions at RHIC energies \cite{sri}. This is expected because 
high mass dileptons are mainly produced in the early stage of
the partonic matter, when the transverse flow has not much developed.

\begin{figure}[ht]
\setlength{\epsfxsize=\textwidth}
\setlength{\epsfysize=0.7\textheight}
\centerline{\epsfbox{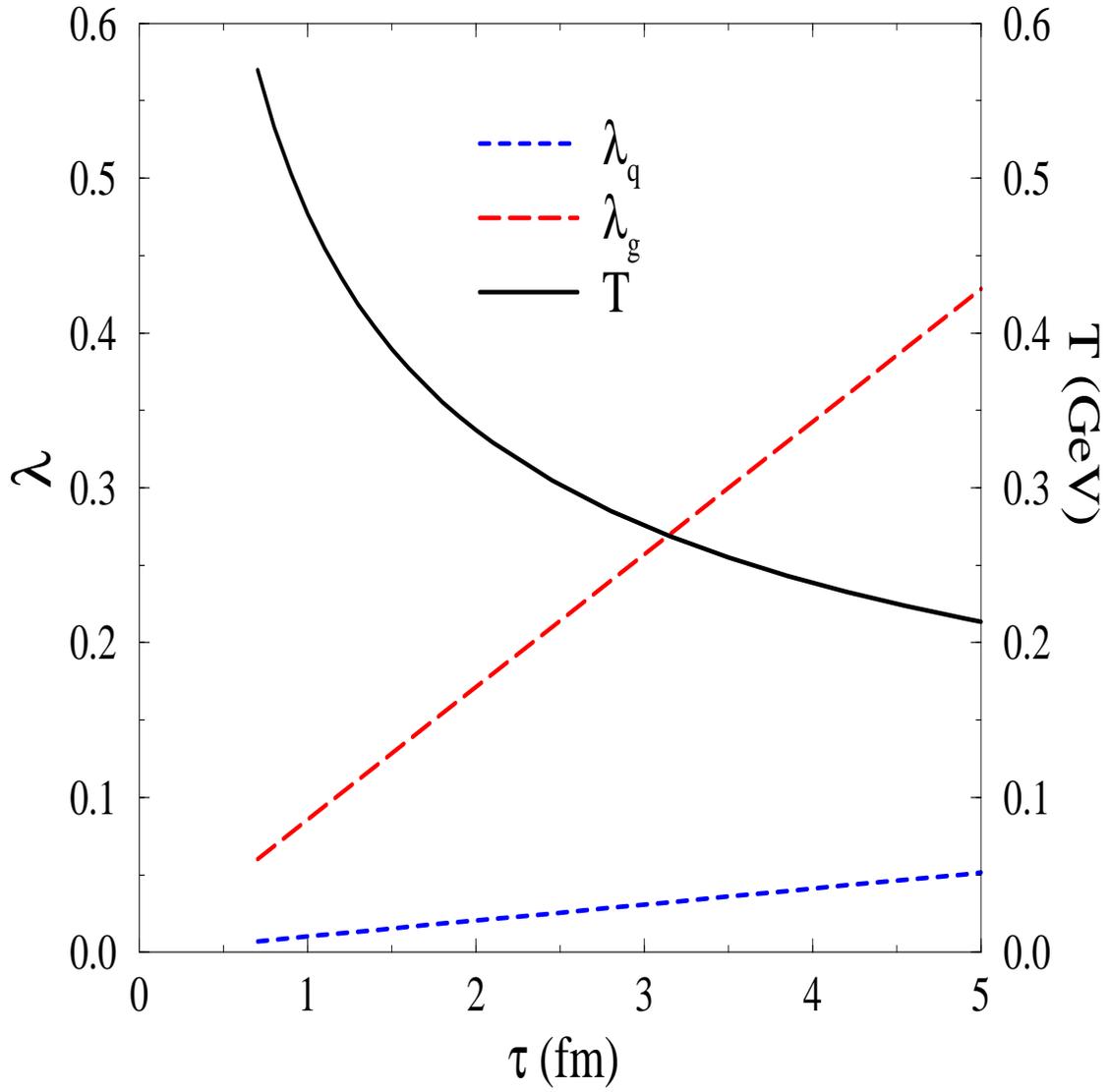}}
\caption{
Time evolution of parton fugacities and temperature.
}
\label{fig_time}
\end{figure}

In Fig. \ref{fig_time} the time evolution of fugacities and
temperature given in Eqs.(\ref{timef}) and (\ref{timet}) are
shown. We see that the temperature decreases faster than that
given by the scenario of entropy conservation, i.e., $T \propto
1/\tau^{1/3}$, as parton production increases the entropy.

\subsection{The default scenario}
\label{subsec-lower}

As in Ref. \cite{xn-pr}, where the parton equilibration is studied
using kinetic equations, we assume that at the proper time
$\tau_0$=0.7 fm/$c$, thermal equilibrium is established at a
temperature $T_0=570$ MeV. We also take the same gluon and quark
densities at $\tau_0$ as in Ref. \cite{xn-pr}, and obtain the
following values for the initial parton fugacities \cite{fugacity}:
\begin{eqnarray}
\lambda_{g,0}=0.060{\;\rm
and\;}\lambda_{q,0}=0.0072.
\end{eqnarray}
With these values, we find that quarks and antiquarks make about
15\% of the partons and the thermal quark mass is about 75 MeV if
the strong coupling constant is evaluated at the normalization
scale of 2 GeV. Subsequent expansion of the partonic matter is
assumed to follow the Bjorken scaling.

\begin{figure}[ht]
\setlength{\epsfxsize=\textwidth}
\setlength{\epsfysize=0.7\textheight}
\centerline{\epsfbox{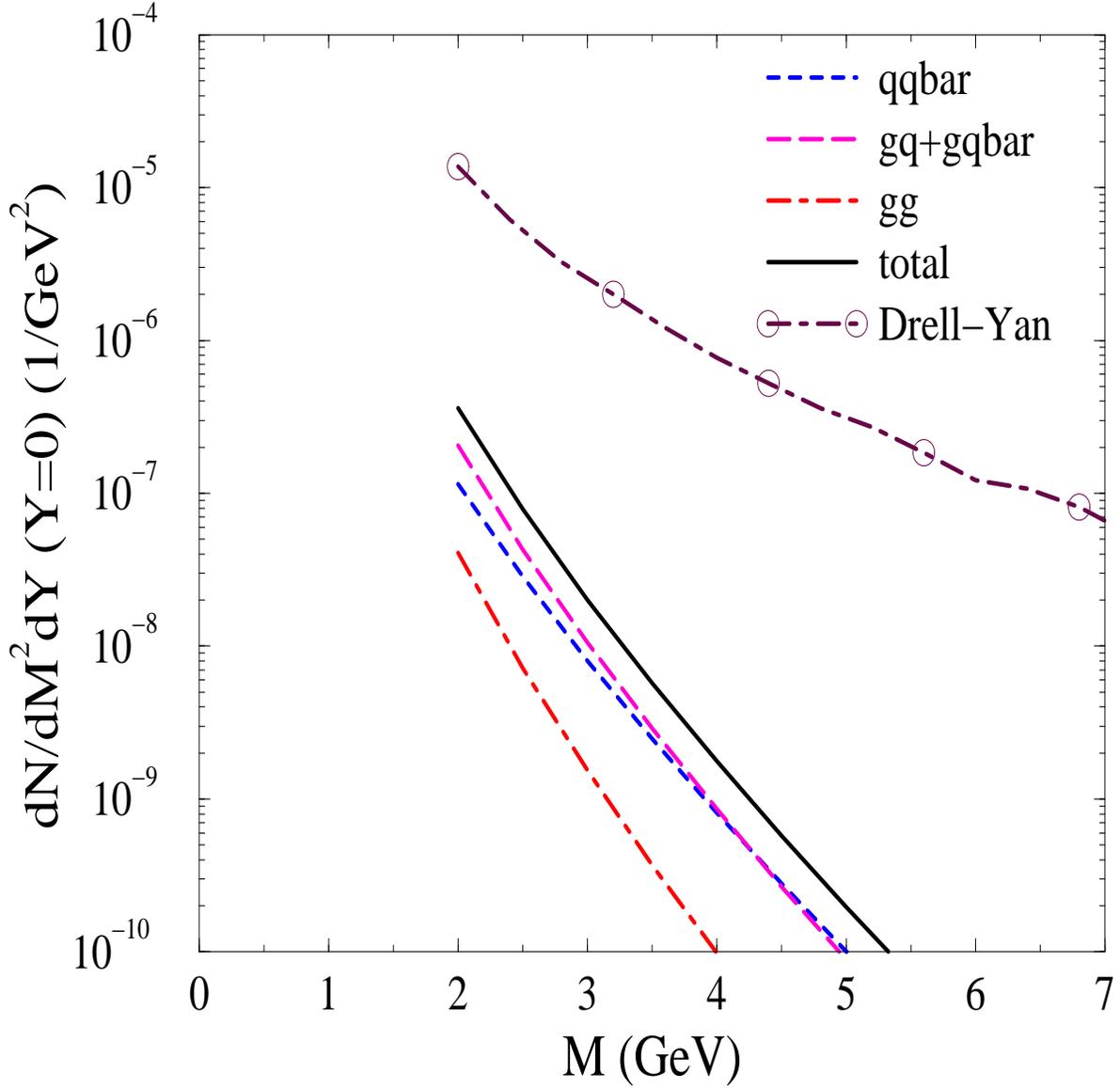}}
\caption{
Thermal dilepton yields for central Au+Au collisions 
at RHIC with initial conditions
of $T_0=570$ MeV, $\lambda_{g,0}=0.060$, $\lambda_{q,0}=0.0072$
and the time evolution according to Eqs.(\ref{timef}) and (\ref{timet}).
}
\label{fig_default}
\end{figure}

We have evaluated the thermal dilepton yield using Eq.(\ref{dndm2dy}), 
and the results are shown in Fig. \ref{fig_default}. Also shown is 
the yield from primary Drell-Yan hard collisions using a 
K-factor of $1.7$, obtained from fitting the E772 data, to take into 
account higher-order contributions. As in thermal dilepton rates shown 
in Fig.\ref{fig_rate}, we find that the Compton processes 
produce more dileptons than both the LO $q\bar q$ annihilation 
and the two-gluon fusion processes. 
This also indicates that the conventional way of using a K factor of 2 
to account for all higher-order contributions \cite{strickland} 
underestimates the possible large enhancement from gluonic processes. 
We note that, despite the enhancement from gluonic processes, 
the total thermal dilepton yield is still a factor of 30 or more 
(depending on the dilepton mass) below the Drell-Yan yield. It is 
thus unlikely that thermal dileptons from the partonic matter can
be observed in heavy ion collisions at RHIC.

\subsection{An upper limit}
\label{subsec-upper}

The small thermal dilepton yield discussed in the above default 
scenario is mainly due to the small number of quarks and antiquarks 
compared to that of gluons assumed in the initial conditions for
the partonic matter. While we have used the ratio of initial quarks 
to gluons that is consistent with minijet calculations based on perturbative 
QCD together with initial and final state radiations, it is possible 
that additional soft partons may be produced. Also, there may be 
already some chemical equilibration before the thermal stage starts 
at $\tau_0$. Both could then lead to initial quark fugacity that 
is larger than the estimate from minijet calculations.

To obtain an upper limit for the thermal dilepton yield, we assume
that quarks and antiquarks are in relative chemical equilibrium
with gluons at the initial time $\tau_0$, i.e.,
$\lambda_q=\lambda_g$, although both are still out of full
chemical equilibrium. Requiring the same value of total $dE_{\rm
T}/dy$ at central rapidity as before and using the same $\tau_0$ and $T_0$, 
we obtain from Eq.(\ref{detdy}) 
the following initial parton fugacities for the upper limit (UL):
\begin{eqnarray}
\lambda_{g,0}^{UL}=\lambda_{q,0}^{UL}= \frac{d_g
\lambda_{g,0}+(d_q+d_{\bar q})N_f \lambda_{q,0}} {d_g+(d_q+d_{\bar
q})N_f} \simeq 0.028.
\end{eqnarray}
This gives a ratio of 3/5 for the sum of quarks and antiquarks to 
the total number of partons, similar to that in an ideal gas of 
gluons and quarks with $N_f=2$.

\begin{figure}[ht]
\setlength{\epsfxsize=\textwidth}
\setlength{\epsfysize=0.7\textheight}
\centerline{\epsfbox{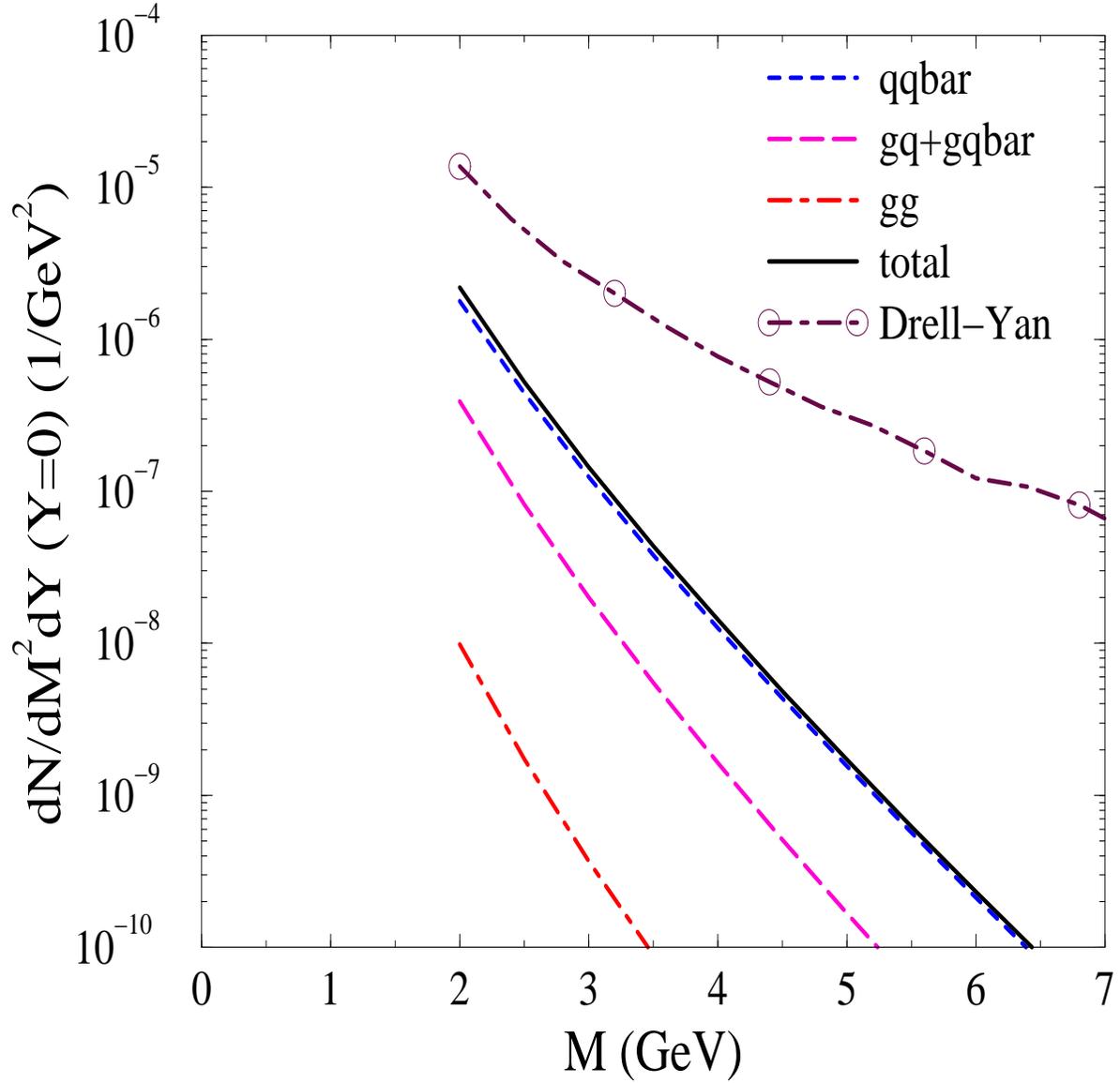}}
\caption{
Upper limit of thermal dilepton yields for central Au+Au collisions at RHIC
with initial conditions of $T_0=570$ MeV
and $\lambda_{g,0}=\lambda_{q,0}=0.028$.
}
\label{fig_equal}
\end{figure}

Using the same time evolution as given by Eqs.(\ref{timef}) and
(\ref{timet}), we have evaluated the thermal dilepton yield with
the above initial parton conditions, and the results are shown in
Fig.\ref{fig_equal}.  Because of increased number
of quarks and antiquarks, the contribution from the LO $q\bar q$
process is increased by a factor of about $16$, while gluonic
processes become less important in comparison. As a result, the total thermal
dilepton yield is about a factor of six higher than that in the default 
scenario. Depending on the dilepton mass, this is still
about a factor of $5$ or more below the Drell-Yan yield.

\subsection{Dependence on initial parton conditions}

The thermal dilepton yield is sensitive to the initial conditions, 
such as $\tau_0$ and $T_0$, for the partonic matter.
For example, the contribution from $q\bar q$ annihilation given in 
Eq.(\ref{rate2}) depends on temperature as
\begin{eqnarray}
\frac{dR_{q\bar q}}{dM}\propto T^{3/2} e^{-M/T}. 
\label{t0dep}
\end{eqnarray} 
The thermal dilepton yield can be large if the partonic
matter starts at a very high initial temperature $T_0$. 
In the following, we study the dependence of the thermal dilepton
yield on the parameters $\lambda_{g,0}$, $\lambda_{q,0}$, $\tau_0$
and $T_0$.

For a fully-equilibrated QGP at $T_0=570$ MeV with $N_f=2$ and an
initial proper time $\tau_0=0.7$ fm/$c$, the rapidity density of
transverse energy is \footnote{Bose-Einstein and Fermi-Dirac
distributions are used.}
\begin{eqnarray}
\frac{dE_{\rm T}}{dy} =\tau_0 \pi
R_A^2 \frac {\pi}{4} \left (\frac{8}{15}+\frac{7N_f}{20}\right )
\pi^2 T_0^4 \sim 12,000 {\rm\; GeV}
\end{eqnarray}
at central rapidity. This is more than an order of magnitude
larger than most model estimates of the final $dE_{\rm T}/dy$.

On the other hand, Eq.(\ref{dndy}) with $\lambda_{g,0}=0.060$,
$\lambda_{q,0}=0.0072$, $\tau_0=0.7$ fm/$c$ and $T_0=570$ MeV
in the default scenario gives at central rapidity
\begin{eqnarray}
\frac{dE_{\rm T}}{dy} =
\frac{dN}{dy} \frac{3\pi T}{4} \sim 350 {\rm \;GeV}.
\label{et}
\end{eqnarray}
Including additional $E_{\rm T}$ from subsequent string
fragmentation gives a final $dE^F_{\rm T}/dy \sim 900$ GeV at
central rapidity, which is comparable with predictions from most
event generators and cascade models for heavy ion collisions at
RHIC energies.

In the following, we shall study the dependence of thermal
dilepton yields on the initial conditions that have same value of
initial $dE_{\rm T}/dy$ at central rapidity. According to
Eq.(\ref{detdy}), this is equivalent to taking initial conditions
with the constraint
\begin{eqnarray} 
\left ( \sum_i d_i \lambda_{i,0} \right )
\tau_0 T_0^4={\rm const}. \label{constraint}
\end{eqnarray}
We have already studied the effect due to the increase of the
quark fraction in the partonic matter. Here, we study the
dependence on initial conditions with a fixed
$\lambda_{g,0}/\lambda_{q,0}$, i.e., with same values of both
$\lambda_{g,0} \tau_0 T_0^4$ and $\lambda_{q,0} \tau_0 T_0^4$.

Using the same time evolution given by Eqs.(\ref{timef}) and (\ref{timet}),
the LO $q\bar q$ annihilation process gives
\begin{eqnarray}
\frac{dN_{q\bar q}}{dM^2 dY}
&\simeq& \frac{\pi R_A^2}{2M} \int_{\tau_0}^{\tau_f}
d\tau \tau \frac{dR_{q\bar q}}{dM}(\tau) \propto \int_{\tau_0}^{\tau_f}
d\tau \tau \lambda_q^2 T K_1 \left ( \frac{M}{T} \right )\nonumber \\
&\propto& \left ({\lambda_{q,0}} \tau_0 T_0^4 \right )^2
\int_{\frac{M}{T_0}}^{\frac{M}{T_c}} dx  x^6 K_1 (x)
\propto \int_{\frac{M}{T_0}}^{\infty} dx x^6 K_1 (x).
\label{dep}
\end{eqnarray}
It is seen that this contribution depends only on one
dimensionless variable $M/T_0$.

\begin{figure}[ht]
\setlength{\epsfxsize=\textwidth}
\setlength{\epsfysize=0.7\textheight}
\centerline{\epsfbox{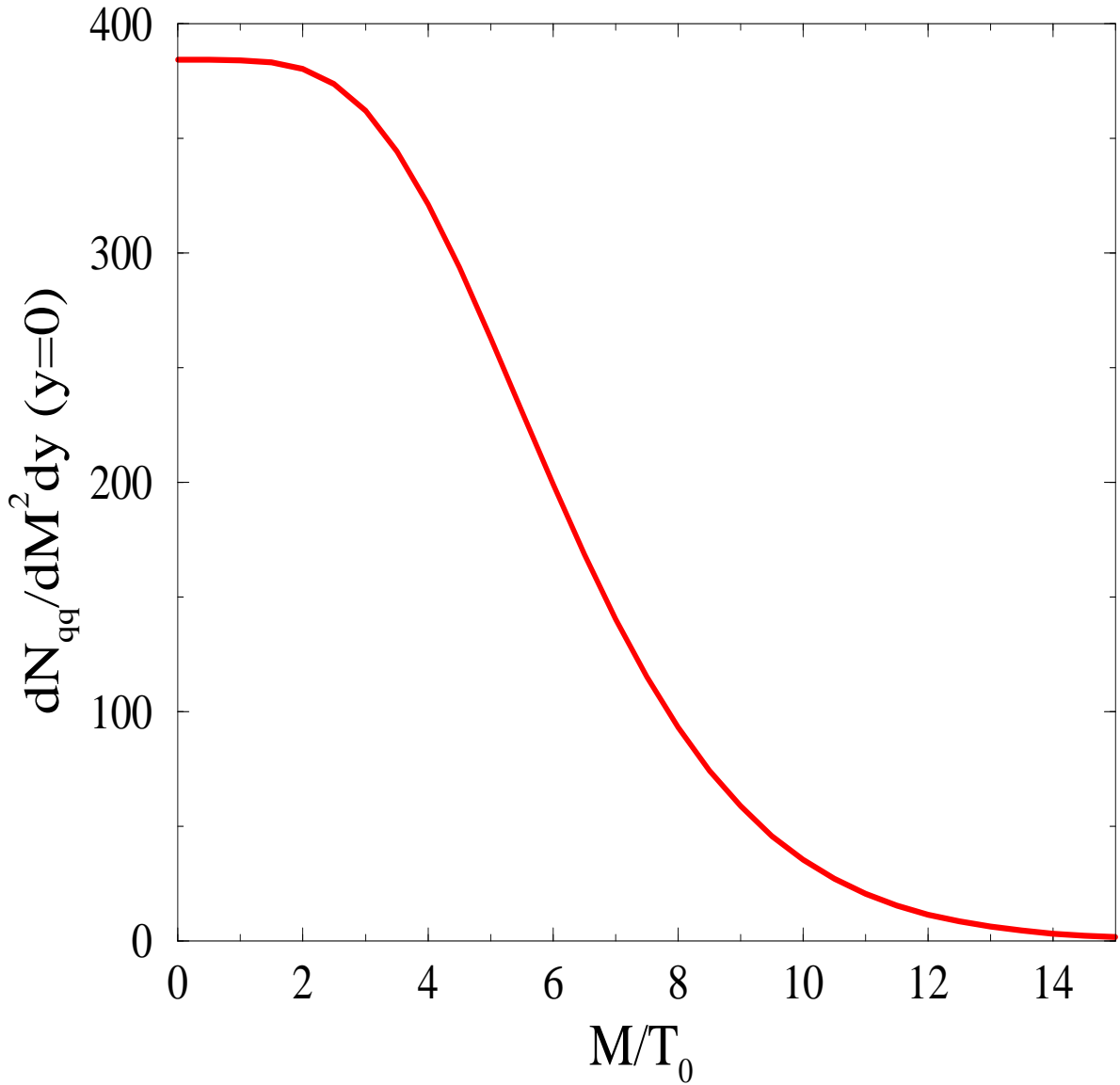}}
\caption{
Dependence of the thermal dilepton yield from the LO $q\bar q$ annihilation
on the initial temperature $T_0$ with fixed
rapidity density of the initial transverse energy. 
See Eq.(\ref{dep}) for the unit of the ordinate.
}
\label{fig_dep}
\end{figure}

In Fig.\ref{fig_dep}, we show the value of the last integral in the 
above equation as a function of $M/T_0$. As $T_0$ increases, the 
dilepton yield at a fixed mass $M$ increases as the variable $M/T_0$ 
shifts to the left, and the yield saturates at the high temperature 
limit. The slow increase and final saturation with increasing initial 
temperature $T_0$ deviate from the naive expectation based on 
Eq.(\ref{t0dep}). This is due to the constraint on the total 
$dE_{\rm T}/dy$ in Eq.(\ref{constraint}), which requires that as 
the initial temperature increases, the initial volume and/or the initial
parton fugacities become smaller, thus making the total dilepton
yield less sensitive to the initial conditions. We also see that the
yield of higher-mass dileptons depend more strongly on the initial
temperature. However, in mass range of $2-4$ GeV,
where the yield of thermal dileptons is closer to the Drell-Yan
yield, the variable $M/T_0$ has a value between $3.5$ to $7$ when
the default choice of $T_0=570$ MeV is used. Fig.\ref{fig_dep}
thus shows that the maximum increase one can obtain by increasing
$T_0$ while keeping the same total $dE_{\rm T}/dy$ is
only $15\%$ at dilepton mass of $2$ GeV and no more than a factor
of 2 at dilepton mass of $4$ GeV. Therefore, thermal dileptons are
still well below the Drell-Yan yield even if we allow the maximum
increase of the upper limit results in Fig. \ref{fig_equal}. 

Same studies can be carried out for the Compton and two-gluon
processes, but they involve more parameters besides the
dimensionless variable $M/T_0$ due to the presence of the
thermal quark mass $\mu$. Relations similar to Eq.(\ref{dep}) can
still be obtained except for the logarithmic factors shown in
Eq.(\ref{scale}). Similar conclusions thus hold as long as the
screening mass is not extremely small.

\section{Discussions and Summary}
\label{sec-final}

In the present study, only leading-order processes have been considered. 
For low-mass dilepton production, higher-order effects have been studied 
\cite{aurenche} using the Hard-Thermal-Loop (HTL) resummation 
method with effective propagators and vertices. 
It has been found that two-loop diagrams can give contributions at 
the same order as one-loop diagrams.  Numerical calculations
have shown that in heavy ion collisions at RHIC and LHC energies 
the two-loop contribution from the bremsstrahlung processes 
is important only for dileptons with mass $M\ll T$ and becomes less 
important than the leading-order contribution for dileptons with mass 
$M > 0.3$ GeV \cite{pal}. Since our study mainly concerns dileptons with
mass above $2$ GeV ($M\gg T$), we do not expect that the inclusion of 
these processes will significantly change our results.

\begin{figure}[ht]
\setlength{\epsfxsize=\textwidth}
\setlength{\epsfysize=0.7\textheight}
\centerline{\epsfbox{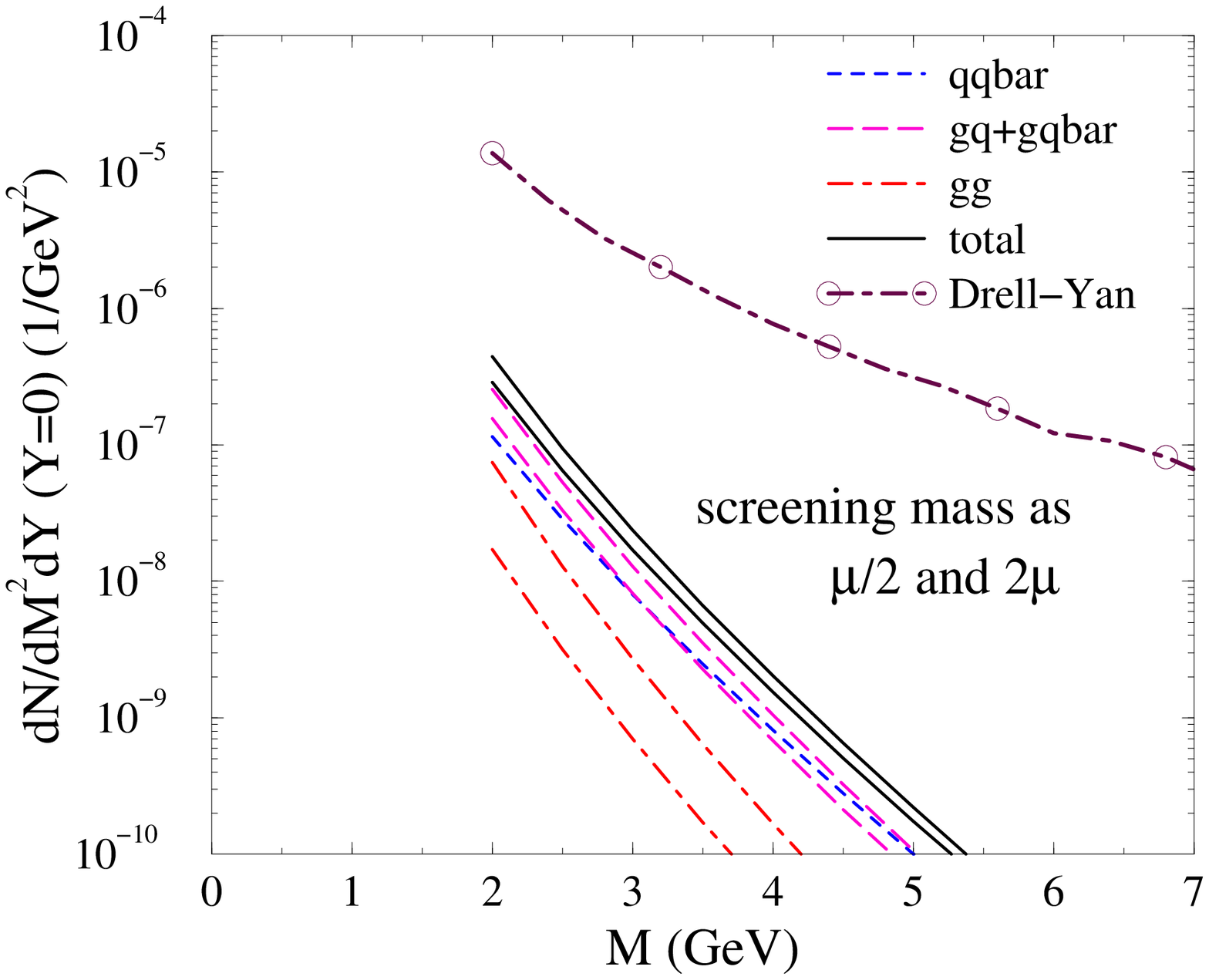}}
\caption{
Screening mass dependence of thermal dilepton yields
for central Au+Au collisions at RHIC with initial conditions of $T_0=570$ MeV,
$\lambda_{g,0}=0.060$ and $\lambda_{q,0}=0.0072$.
Upper and lower curves correspond to results using half and twice
the thermal quark mass as the screening mass, respectively.
}
\label{fig_mu2}
\end{figure}

However, there are still uncertainties in our results.
First of all, the divergence in the dilepton
production cross sections has been simply regularized by the 
thermal quark mass. 
Studies that take into account 
higher-order effects using the thermal field theory have indeed 
shown that for low mass dilepton or photon production in the QGP
the thermal quark mass provides a self-consistent 
screening effect \cite{altherr,thoma,baier2}.  
However, the dependence of the dilepton production cross section on the 
thermal quark mass differs from that based on the lowest-order 
perturbative calculations as used in our study. Also, 
using different parton distributions can lead to different
dependence of the cross sections on the thermal quark mass \cite{baier2}.
Since we have not carried out a self-consistent calculation 
including higher-order effects and have used the Boltzmann 
distribution instead of the quantum distribution, the
coefficients that appear in association with the thermal quark mass
as given in Eqs.(\ref{screen1}) and (\ref{screen2}) 
are expected to be different when a more complete calculation is 
carried out. To illustrate this
uncertainty, we show in Fig.\ref{fig_mu2} the results from changing the 
screening mass by a factor of two but with the same default initial 
parton conditions. Since dilepton yields from the Compton and 
two-gluon processes roughly scale as $\log \mu$ and $\log^2 \mu$, 
respectively, sizable differences are seen. 

We have only considered dilepton production from
the stage where partons have reached thermal equilibrium. The contribution 
from the pre-thermal stage has not been included. Also, we have not 
considered chemical equilibration during the pre-thermal stage, which 
could change the initial conditions of the partonic matter such as 
the fugacities $\lambda_{g,0}$ and $\lambda_{q,0}$. With the 
dilepton production cross sections given in Section
\ref{sec-xsection}, the pre-thermal dilepton yield can be
calculated using a pre-thermal parton distribution that is
either parameterized or from a parton cascade code. With the latter, the
entire time evolution of parton thermal and chemical equilibration can be
studied, so the separation of pre-thermal and thermal stages,
which is rather crude, is unnecessary. In this way, the
evolution of parton fugacities can be self-consistently
determined, and all secondary dileptons from the partonic stage
can be calculated.  In a previous study \cite{qm99}, we have carried 
out a calculation of secondary dileptons from the LO $q\bar q$ annihilation 
process within the parton cascade code ZPC \cite{zpc}, where initial
parton fugacities were determined directly from the HIJING model
\cite{hj}. Since the default HIJING model gives too few quarks and
antiquarks, with the ratio of the sum of quarks and antiquarks to
the total number of partons being only 4.5\% (compared with the
ratio of 15\% in the default scenario of the present study based on
minijet calculations using the perturbative QCD), the thermal dilepton
yield in Ref.\cite{qm99} is lower than the default result in this study
by a factor of 10.

In summary, we have studied thermal dilepton production in heavy ion
collisions at RHIC energies. In particular, contributions from
gluonic processes and their dependence on the initial parton
conditions are investigated. We find that gluonic processes
significantly enhance the thermal dilepton yield as one expects
more gluons than quarks in the partonic matter formed in heavy ion
collisions at RHIC. However, we have also shown that with the 
constraint of Eq.(\ref{et}) on the total initial $dE_{\rm T}/dy$ of 
the partonic matter, thermal dileptons are always significantly lower 
than those from the Drell-Yan process for any reasonable variations 
of the initial temperature, proper time, and fugacities. Our results 
show that thermal dileptons can compete with those from the Drell-Yan 
process only in a rather ideal case, when the initial energy density 
is higher than most current expectations, and parton chemical 
equilibration proceeds fast so that quarks are close to relative 
equilibrium with gluons. 

\bigskip\bigskip
\noindent{\bf Acknowledgments:}
We thank B. K\"ampfer, W.L. van Neerven and S.M.H. Wong 
for helpful discussions. This work was supported in part by
the National Science Foundation under Grant No. PHY-9870038, the
Welch Foundation under Grant No. A-1358, and the Texas Advanced
Research Project FY97-010366-068.

\pagebreak
{}

\pagebreak

\pagebreak

\pagebreak

\pagebreak

\pagebreak

\pagebreak

\pagebreak

\pagebreak

\pagebreak


\begin{thebibliography}{99}
\bibitem{fugacity}
The difference between the values of our parton fugacities 
($\lambda_{g,0}=0.060$, $\lambda_{q,0}=0.0072$) and those in 
Ref.\cite{xn-pr} ($\lambda_g^0=0.05$, $\lambda_q^0=0.008$) 
for the same parton number densities is due to the fact that
we use the factorized Boltzmann distribution instead 
the factorized Bose-Einstein and Fermi-Dirac distributions.
\end{thebibliography}
\end{document}